\documentclass{ws-procs9x6}

\usepackage{epsfig}

\newcommand{\be}{\begin{equation}}
\newcommand{\ee}{\end{equation}}
\newcommand{\bea}{\begin{eqnarray}}
\newcommand{\eea}{\end{eqnarray}}

\begin{document}

\title{Neutrino Physics after KamLAND}

\author{A.~Smirnov 
\footnote{\uppercase{I}nvited talk given at the 
\uppercase{I}nternational workshop
\uppercase{NOON}2003,  \uppercase{F}ebruary 10 - 14,  2003, \uppercase{K}anazawa,  
\uppercase{J}apan. (\uppercase{C}omplete version of the paper.)}}
\address{International Centre for Theoretical Physics\\
Strada Costiera 11 , 34 014 Trieste, Italy,\\ 
Institute for Nuclear Research, RAS, Moscow, Russia\\ 
E-mail: smirnov@ictp.trieste.it}


\maketitle

\abstracts{
The neutrino anomalies  were driving force of  
the developments in  neutrino physics during the last 30 - 35 years. 
I will consider  status of the anomalies after 
the first KamLAND result. The main questions  are 
``What is left?" and ``What is the next?"    
In the new phase, the  phenomenological 
objectives of neutrino physics consist of  accomplishing the program of 
reconstruction of the neutrino mass and flavor spectrum  and  searches 
for physics beyond the ``standard'' picture. The latter includes searches for  
new (sterile) neutrino states, new neutrino interactions, 
effects of violation of the fundamental symmetries in the neutrino sector. 
}



\section{Introduction}

\subsection{Before and After}

``After KamLAND'' means essentially after confirmation of the 
large mixing MSW (LMA) solution 
of the solar neutrino problem. 
In this sense,  neutrino physics ``after KamLAND''  has started much 
before the announcement of the first KamLAND result~\cite{KL}. Since  1999 
most of the papers (on theoretical implications and  
phenomenology including physics of the long-baseline experiments) 
has been written in the context of  LMA solution. 

1998 was the turn point.  
The essence  of ``Revolution-98'' inspired by the
SuperKamiokande results on the atmospheric~\cite{SK} and solar~\cite{SKs} neutrinos   
consisted of  

$\bullet$ strong evidence of the  $\nu_{\mu} - \nu_{\tau}$  oscillations of the atmospheric neutrinos
with maximal or nearly maximal mixing,

$\bullet$ strong evidence against the small mixing  MSW solution of the solar neutrino problem. 

The prejudice of small mixing, which was the  dominating idea during many
years, has been destroyed.

Already in 1998, the solar neutrino data gave some hint that  the large mixing MSW  
effect can be the solution of the solar neutrino problem \cite{BKS}.  
With more data appeared,   LMA  became favored and then    
the most plausible explanation.  
On the basis of LMA, detailed  predictions for KamLAND  
have been done \cite{ped}.

The KamLAND result is the  culmination of about 40 years 
of the solar neutrino studies. 
This result is the  confirmation of not only LMA (in assumption of the CPT symmetry), 
but also the whole oscillation picture
behind the neutrino anomalies including the oscillations of 
atmospheric neutrinos.

\subsection{The end of era of the neutrino anomalies?}

Neutrino anomalies, both real and fake, were driving force of developments 
in the field.

The  famous triplet is  solar-atmospheric-LSND.
The atmospheric neutrino anomaly
and  the solar neutrino problem turned out to be real (not related to 
experimental or systematic errors), confirmed and practically 
resolved.  The  LSND anomaly~\cite{LSND} is badly resolved,     
not confirmed,  but not yet excluded.

Fake anomalies played  certain positive role
attracting the interest to the field, forcing to think and 
... invent sometimes correct theoretical ideas ({\it e.g.} neutrino oscillations).
The list includes the  17 eV (ITEP) mass, 17 kev neutrino,
BUGEY oscillations, KARMEN anomaly, Troitzk anomaly,  {\it etc.}.

There are some problems in physics and  astrophysics which may be 
related to neutrinos:  Ultra-high energy cosmic rays beyond the GZK limit, 
nucleosynthesis of heavy elements, large pulsar kicks, {\it etc.}.

The main  questions  now are ``What is left?'' and ``What is the next?'' 

\section{Is the solar neutrino problem solved?}

\subsection{Solar neutrinos and KamLAND}

\begin{figure}[h!]
\begin{center}
\vspace{1.5cm}
\begin{tabular}{cc}
\hspace{-1cm} \epsfxsize6.5cm\epsffile{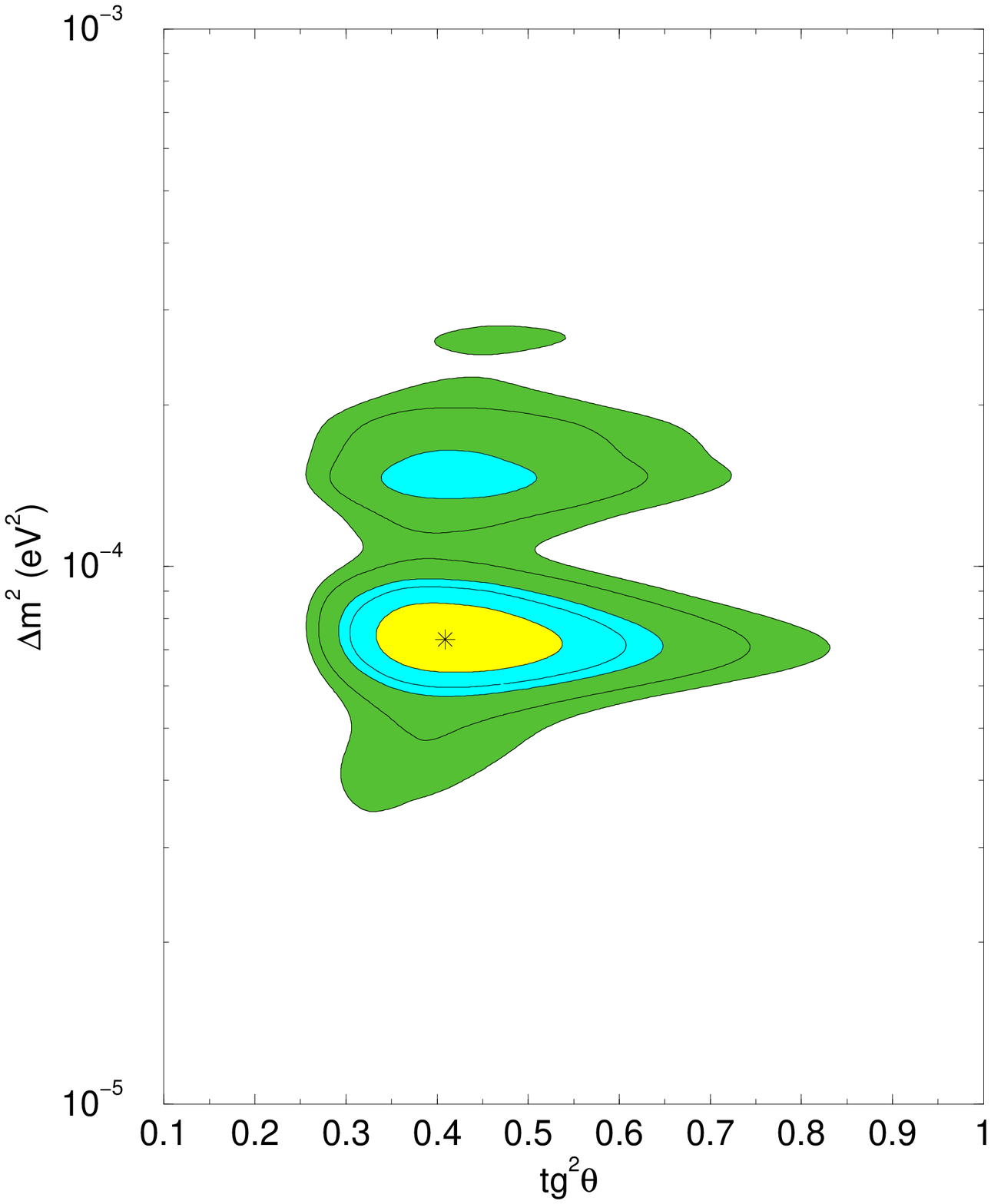} & \hskip -1cm
              \epsfxsize6.5cm\epsffile{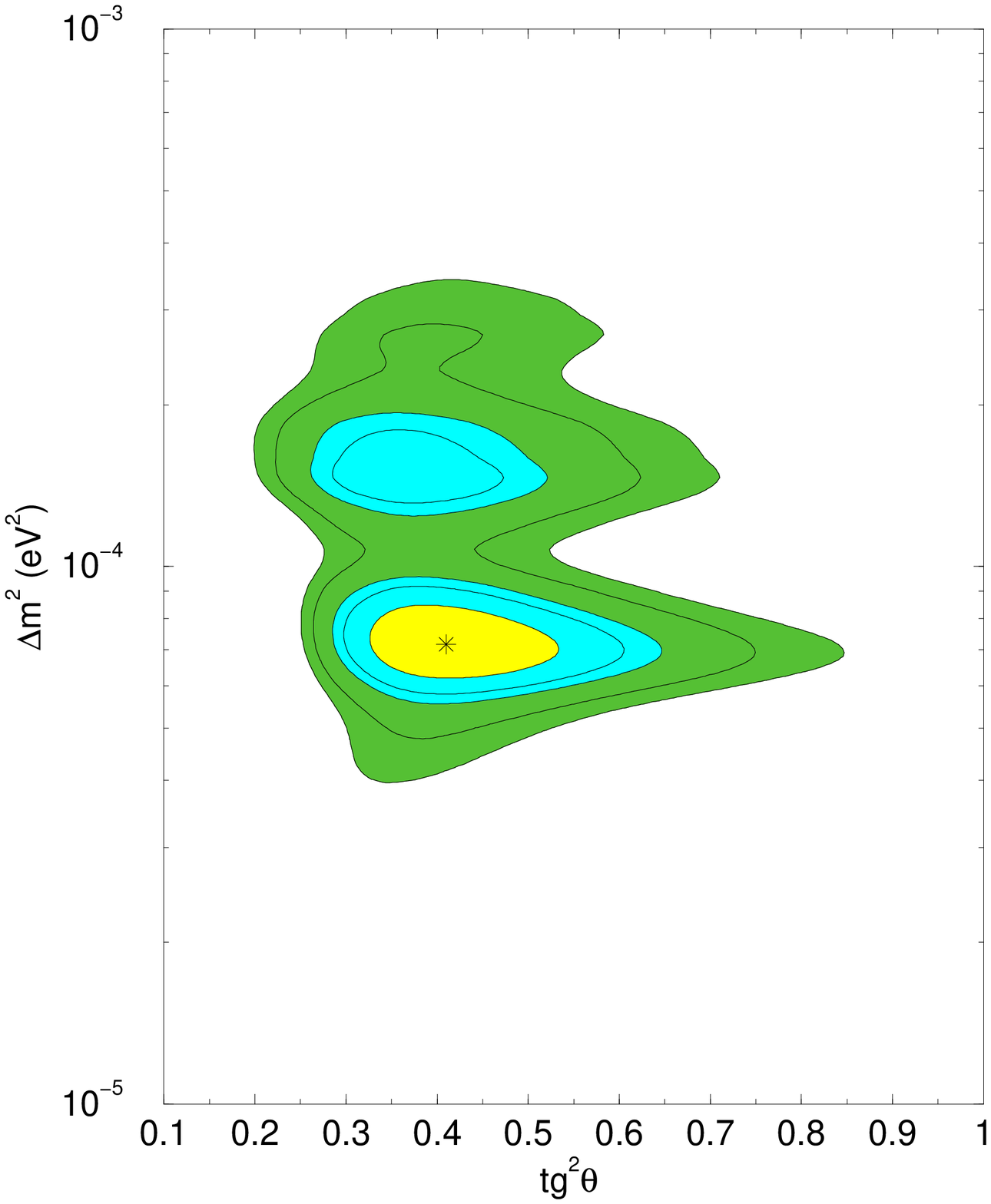}
\end{tabular}
\vspace{-1cm}
\caption{\it
The allowed   regions  of oscillation parameters from
the combined analysis of the solar neutrino data and the KamLAND spectrum at 
$1\sigma$ (inner region), $90\%$, $95\%$, $99\%$ and $3\sigma$  C.L..
The left panel: $s_{13} = 0$,  the right panel: $s_{13} = 0.2$. 
From $^8$.}
\label{sol}
\end{center}
\end{figure}

The first KamLAND result (see analysis in\cite{solar,pedKL} and 
fig.~\ref{sol} from\cite{pedKL}) 

$\bullet$  has confirmed (in assumption of CPT) the LMA MSW solution and 
excluded other suggested effects at least as the dominant mechanisms. 

$\bullet$ has further shifted the allowed region  and the best fit point to larger 
values of $\Delta m^2$:
\be
\Delta m^2  = (5 \rightarrow 7)\cdot 10^{-5}~ {\rm eV}^2, 
\ee

$\bullet$ put the  lower bound on $\Delta m^2$ 
\be
\Delta m^2 > (4 - 5) \cdot 10^{-5} {\rm eV}^2, 
\ee
which looks rather solid:  for smaller $\Delta m^2$  the strong 
distortion of the spectrum is predicted which contradicts the data. 

\begin{figure}[h!]
\begin{center}
\vspace{1cm}
\hspace{-0.1cm} \epsfxsize7cm\epsffile{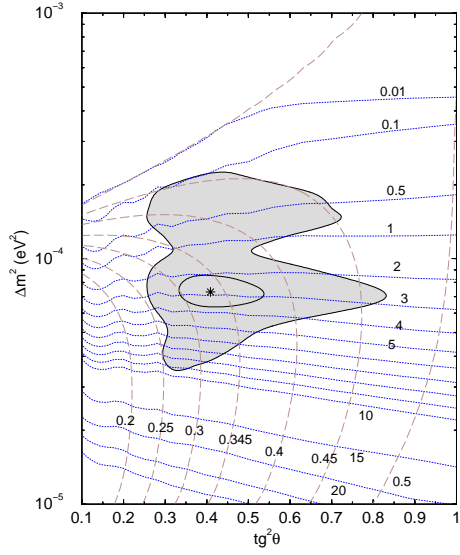}
\vspace{-1cm}
\caption{\it The allowed $1\sigma$ and $3\sigma$ regions (shadowed) of oscillation parameters from
the combined analysis of the solar neutrino data and KamLAND. 
Shown are also contours of the constant Day-Night asymmetry (dotted lines) and 
the CC/NC ratio (dashed lines). From $^8$.}
\label{fit} 
\end{center}
\end{figure}

In fig.~\ref{sol} I show the allowed regions of oscillation parameters from 
the combined analysis of the solar neutrino data and KamLAND~\cite{pedKL}. 
Clearly, there is no enough sensitivity to $s_{13}$ at present.

\subsection{LMA: precision measurements}

Further decrease of  the allowed  $\Delta m^2 - \tan^2\theta$ region 
is needed for a number of  reasons: for theoretical implications, 
further phenomenological and experimental developments, and  also  
for precise understanding the  physical picture of 
neutrino conversion. Indeed, 
high values of $\Delta m^2$ correspond to vacuum oscillations 
with small matter corrections. 
Low values of $\Delta m^2$
- to the non-oscillatory adiabatic conversion (for $E > 5$ MeV). 
Small mixing admits clear resonance description.  
Maximal mixing means that the resonance is at zero density. 


The forthcoming improvements are expected  from further operation of SNO and KamLAND. 
In the fig.~\ref{fit} we show the contours of constant ratio CC/NC 
and Day-Night asymmetry  in the plane of oscillation 
parameters.   
The expected accuracy of the  measurements of CC/NC is 
about 10\%,  so that the two regions (h- higher  and l - lower) can be  
distinguished. In the range 
$\Delta m^2 < 10^{-4}$ eV$^2$, SNO will give 
preciser determination of the mixing angle and more 
stringent bound on the deviation from maximal mixing. 
Later KamLAND will achieve 10\% accuracy in $\Delta m^2$. 
This accuracy is comparable with a possible effect of $s_{13}$. 

In future the  analysis of data will be performed in two stages:

1). Identification of the unique region (discrimination between
LMA-h and LMA-l). At this stage the analysis of data can be performed
in the context of two neutrino mixing and the sub-leading effects due
to the 1-3 mixing can be neglected.

2). Precision measurements. Possible sub-leading
effects should be included. The generic $3\nu$- analysis should be
performed. Problem of degeneracy of parameters will appear.

\subsection{Consistency checks}

Till now no a single signature of the LMA solution 
({\it e.g.},  the day-night asymmetry, upturn of the spectrum)
has been observed at a statistically significant level.  
What is expected? The following predictions  correspond to the present 
best fit region~\cite{pedKL}: 

$\bullet$ the Day-Night asymmetry at SNO and SK: 
\be
A_{DN} (SNO) = (2 - 5) \% ,  ~~~~~A_{DN}(SK) = (1 - 3) \% , 
\ee

$\bullet$ spectrum distortion: 
the 5 - 10 \% upturn  is expected at low energies between 8 and 5 MeV,  

$\bullet$ suppression of the signal at the intermediate energies 
(BOREXINO and KamLAND): 
\be
R_{B} = (0.6 - 0.65) R_{B}^{SSM},  
\ee
(where the effect of the NC interactions 
for $\nu - e$ scattering  has been taken into account).  

$\bullet$  small seasonal variations: 
the expected winter-summer asymmetry of the signal at SNO and SK,   
$
A_{WS} < 0.5 \%
$, 
is practically unobservable,  

$\bullet$ suppression of the $pp$- neutrino flux:  
$R_{pp} = 0.6$
which can be measured in future low energy solar neutrino experiments.

Tests of these predictions have the  threefold implication: 
(i) further confirmation of LMA: one needs to over determine the solution 
to perform its cross-checks;  
(ii) precise determination of the neutrino parameters; 
(iii) searches for  physics ``Beyond  LMA''. 

\subsection{Homestake anomaly?}
 
Quality of description of the available data by 
the LMA solution is very good:   
according to the global fit $\chi^2 /d.o.f. = 68.2/91$.
This is also  confirmed by the pull-of diagram.  
A visible deviation appears in one place only:  
LMA predicts about $\sim 2 \sigma$ higher Ar-production 
rate as compared with the Homestake result. 

Possible interpretation? (i) the  statistical fluctuation; 
(ii) unknown systematics, probably related to the claimed time variations 
of the Homestake signal (in some periods of time 
the efficiency of detection was lower);  
(iii) neutrino physics beyond LMA. 

The latter can be related to another observation: the  
absence of apparent upturn of the spectrum (ratio of the observed spectrum 
to the SSM prediction) at low energies.  
Neither SK nor SNO see any upturn, though the sensitivity 
may not be enough. 

Both the lower $Ar$-production rate and the absence (suppression) 
of the  upturn  can be due to the effect of  additional 
(sterile) neutrino  which mixes very weakly with 
active neutrinos (mainly, $\nu_s - \nu_1$) 
and has small  mass split with the lightest state $\nu_1$: 
\be  
\sin^2 2 \theta_{s1} =  (10^{-4} - 10^{-3}), 
 ~~~~\Delta m^2_{01} = (0.5 - 1)\cdot 10^{-5} {\rm  eV}^2.  
\ee
Such a neutrino produces an additional dip in the suppression pit 
in the range 0.8 - 5  MeV,  thus suppressing the Be-neutrino line 
or/and  the upturn of spectrum.  BOREXINO and KamLAND can check this. 

\subsection{Solar neutrinos versus KamLAND}

{}From the  $2\nu$ analysis of the solar neutrino data  
and independent $2\nu$ analysis of the KamLAND results   
one finds that values of parameters in the best fit points   
coincide
\be
(\Delta m^2, \tan^2 \theta)_{solar} \approx 
(\Delta m^2, \tan^2 \theta)_{KamLAND}
\label{equal}
\ee
within $1\sigma$.  This indicates that CPT is conserved in 
the leptonic  sector. 

It is interesting to further check the equality with increasing
accuracy.  The mismatch of parameters 
can testify for the CPT violation or, more probably,  for certain  physics beyond LMA. 

If some effect influences the KamLAND signal it should also 
show up in the solar neutrinos. 
Inverse is not true: a number of  effects can influence the solar 
neutrinos but not KamLAND. 
The solar neutrinos 
have much higher sensitivity to physics beyond  LMA than KamLAND. 
Some  examples: an  additional 
neutrino state with small  
$\Delta m^2$ or/and $\tan^2 \theta$,  
the neutrino spin-flip in the Sun, 
non-standard interactions of neutrinos.  

There is another interesting aspect of comparison of  the 
oscillation parameters extracted from the solar neutrinos  
and KamLAND, namely, - a test of  theory of the conversion and oscillations. 
Indeed, physics behind the solar neutrino conversion and 
the oscillations of reactor neutrinos is different. 
In the case of solar neutrinos we deal with the adiabatic conversion;  
the matter effect dominates (at least in the high energy part of 
the spectrum), the oscillation phase is irrelevant. 
The  effect is described by the adiabatic conversion 
formula. In contrast, in the case of KamLAND, the vacuum oscillations 
occur; the matter effect is very small; the oscillation phase 
is crucial.  Here we use the vacuum oscillation formula.

The coincidence of parameters (\ref{equal})
testifies for correctness of the theory (phase of oscillations, matter potential, 
etc..), see also discussion in~\cite{theor}. 

\subsection{Beyond LMA  or Physics of sub-leading effects}

The physics of sub-leading effects becomes one of the main 
subjects of studies.  The name of the game is 
`` LMA + something'', where LMA provides the leading effect,   
and ``something'' can be $U_{e3}$, 
SFP (spin-flavor precession),  new neutrino states, 
NSI (non-standard interactions),   
VEP (violation of the equivalence principle), {\it etc.}. 
The implications of these studies include the neutrino properties ({\it e.g.}, magnetic moments), physics 
beyond the Standard Model, characteristics of  the interior of the Sun.

``LMA + sterile neutrinos": the  signatures are the modification of the CC/NC
ratio and  additional distortion of the energy spectrum.
An  example has been described in sect. 2.4.

``LMA + NSI": an additional contribution to the matter effect appears;
some deviations from usual relations between
$\Delta m^2$ and $\tan^2 \theta$ are expected.
Some work in this direction has already been performed before
KamLAND announcement~\cite{nun,theor}.

In what follows we will comment on the ``LMA + SFP" scenario. 
If no new neutrino states exist, the only relevant mass difference 
is $\Delta m^2_{LMA}$.  
For such a large $\Delta m^2$ the spin-flip occurs 
in the central regions of the sun (radiative zone) where 
the potential $V \sim \Delta m^2_{LMA}/2E$.  
The signature  of the scenario is the 
appearance of the antineutrino flux.  
For the Boron neutrinos the ratio of $\bar{\nu}_e$- flux to the  
original ${\nu}_e$- flux equals~\cite{AP}  
\be
\frac{\bar F_B}{F_B} = 1.5 \% 
\left(\frac{\mu_{\nu}}{10^{-12}\mu_B} \right)^2 
\left( \frac{B}{100 MG}\right)^2 , 
\label{muB}
\ee
where $\mu_{\nu}$  is the magnetic moment of neutrino, 
$\mu_B$ is the Bohr magneton. 
For $B =7$ MG, that is, at the level of present upper 
bounds~\cite{FG} and  $\mu_{\nu} = 10^{-12} \mu_B$ we get 
$\bar{F}_B /F_B = 7 \cdot 10^{-3} \% $, which is  2 orders of magnitude below the 
present limit~\cite{nubar}.

Unless Voloshin's cancellation~\cite{voloshin} or
polarization suppression~\cite{zee}  in the mass term $m_{\nu}$ occurs,  the
relation exists
\be
\mu_{\nu} \sim \frac{e}{ \Lambda^2} m_{\nu},
\ee
where $e$ is the electric charge and $\Lambda$ is the
energy scale (energy cut) at which the magnetic moment
is formed. For  $\Lambda = 100$ GeV and
$m_{\nu} = 1$ eV we find  $\mu_{\nu} = 10^{-16}\mu_B$
which is practically unobservable. In particular, according to (\ref{muB})
the antineutrino flux is smaller than $10^{-8} \%$.

The spin-flip effect can be much larger, if  new neutrino states 
exist with $\Delta m^2 \ll \Delta m^2_{LMA}$. 

\subsection{Solar neutrino astrophysics}

After resolution of the solar neutrino problem we can come back to the
original task: the spectroscopy of solar neutrinos for studies of  interior
properties of the Sun, that is, neutrino diagnostics of the Sun
\cite{bkz}.
This program  includes determination  of  the original neutrino fluxes, 
in particular, neutrino fluxes from  the pp- and CNO cycles \cite{bah} and 
searches for time variations of fluxes.

Interesting to note that the original $\nu_e$
fluxes as they appear in the SSM do not exist in nature!
According to LMA,  the flavor conversion/oscillations starts
already in the neutrino production region. Still we can
introduce these fluxes since there is  no back
influence of the neutrino conversion on the solar characteristics
and production of neutrinos. Conversion effects can be subtracted.
In any case we can speak on the total flux without specification
of flavor.

\section{Atmospheric neutrinos: Any problem?}

There is a compelling evidence that the $\nu_{\mu} - \nu_{\tau}$ vacuum oscillations 
are  the dominant mechanism of the 
atmospheric neutrino transformations. This evidence is provided 
by SuperKamiokande~\cite{SKa}, MACRO~\cite{MACRO}, SOUDAN~\cite{SOUDAN}. It is confirmed 
by  K2K~\cite{K2K}. 
Nothing statistically significant beyond this interpretation  
has been found so far. 
Though  there is, probably,  some tension between the observed 
up-down asymmetry  of the $\mu$ -like events and 
the  ratio of total numbers of the $\mu$-like and $e$-like events~\cite{los}. 
Also situation with the original neutrino fluxes is not yet clear. 
Eventually this may lead 
to some change of the extracted values of oscillation 
parameters or to discovery of some new sub-leading effects,  
but it will hardly influence the whole oscillation interpretation.

What is the next? As in the case of solar neutrinos: 
the next is physics of the  sub-leading effects and also  geo- and cosmo- physics. 
Among objectives are  searches for 
(i) the $\nu_e$- and $\bar \nu_e$ oscillations, 
(ii) effects of $s_{13}$, 
(iii) sterile neutrinos, 
(iv) CP-violation, 
(v) CPT-violation, 
(vi) specific oscillation effects related to the 
earth density profile (parametric enhancement of oscillations 
for core - mantle crossing trajectories), {\it etc.}.  

Precision measurements of the atmospheric neutrino
fluxes can be used for studies of the cosmic ray fluxes once oscillation
effects are well understood.

\subsection{Oscillations of Atmospheric $\nu_e$}

After KamLAND  we can say 
that  even for $s_{13} = 0$ the $\nu_e$-, $\bar \nu_e$- oscillation effects must appear 
at some level  due to the solar/KamLAND  oscillation parameters. The signature of these 
oscillations is  an excess (or deficit) of the $e$-like events in certain energy range 
with specific energy and zenith angle dependences. 

The $e$-like event excess (deficit)
can  be due to 

(i). oscillations driven by the solar $\Delta m^2_{12}, ~~\theta_{12}$, 
(mainly in the sub-GeV region), 

(ii). oscillations driven by  nonzero  $\theta_{13}$ and atmospheric $\Delta m^2_{13}$  
(mainly in the multi-GeV region), 

(iii). interference of the above effects. 

We will consider these possibilities in order. 

\subsection{Oscillations due to $\Delta m^2_{12},~\theta_{12}$}

A relative change of the $\nu_e$- flux 
due to the oscillations equals~\cite{PS-L} 
\be
\frac{F_e}{F_e^0} - 1 = P_2 (r \cos^2 \theta_{23} - 1), 
\label{sub-e}
\ee
where $P_2 = P(\Delta m^2_{12},~\theta_{12})$ is the $2\nu$ 
transition probability $\nu_e \rightarrow \nu_{\mu, \tau}$ in the matter 
of the Earth, and 
$
r \equiv {F_{\mu}^0}/{F_e^0}. 
$ 
The effect is strongly suppressed by the ``screening factor'' 
(in the brackets of (\ref{sub-e})) in spite of  large 
transition probability, $P_2$. Indeed, in the sub-GeV region 
$r \approx 2$ and the oscillation  effect is zero for 
the maximal 2-3 mixing. This feature  can be used to search for deviations 
of the 2-3 mixing from the maximal one.  The excess (deficit) of the events 
is directly proportional to this deviation. 
In the Table we show the excess 
\be
\epsilon \equiv {N_e}/{N_e^0} - 1
\ee 
in the sub-GeV region 
for different values of $\sin^2 2\theta_{23}$
and for the best fit points of the LMA-l and LMA-h regions.  
\begin{table}
\caption[]{last two columns.}
\begin{tabular}{ccc}
\hline
$\sin^2 2\theta_{23}$ &  $\epsilon_l$, \%  & $\epsilon_h$, \%  \\
\hline
0.91                  &  2.8  &   4.8 \\  
0.96                  &  1.9   &  3.2  \\ 
0.99                  &  0.9   &  1.6  \\   
\hline
\end{tabular}
\end{table}
The excess increases with $\Delta m^2_{12}$, but  does not 
exceed $5\%$. 
Once the LMA parameters are well known, measurements of $\epsilon$ 
will allow to restrict a deviation of the 2-3 mixing  
from the maximal one. 
To realize this  one needs high statistics experiment and 
a possibility to identify the excess due to oscillations  
(in particular,  distinguish it from uncertainties in the normalization
of the original flux).

\subsection{ Effect of  $s_{13}$}

The oscillations driven by  non-zero $s_{13}$ and 
the atmospheric $\Delta m^2_{13}$ produce 
significant modification of the $\nu_e$- flux 
in the multi-GeV region,  where 
the oscillations are enhanced by the matter 
effects (MSW resonances in the mantle and core, 
the parametric enhancement of oscillations)~\cite{ADLS}. 

In contrast to the  sub-GeV sample,  at higher energies $r$ 
significantly deviates from 2 and the screening  
$ \propto (r \sin^2 \theta_{23} - 1)$ is weaker. 
The oscillation effect can be identified by the 
up-down asymmetry: 
\be
A_{U/D} = 2 \frac{U - D}{U + D}, 
\ee
where $U$  and $D$ are the numbers of $e$-like events in the intervals 
of zenith angles
$\cos \Theta = (-1 \div -0.6)$ and $\cos \Theta = (0.6 \div 1)$ 
correspondingly. 
For $\sin^2 2\theta_{23} = 1$  and $\Delta m^2_{13} = (2 - 3) \cdot 10^{-3}$ 
eV$^2$ the asymmetry in the multi-GeV region 
can reach (5 - 8) \% for the maximally allowed 1-3 mixing.

\subsection{Induced interference}

\begin{figure}[h!]
\begin{center}
\vspace{-0.5cm}
\hspace{+1cm} \epsfxsize8.5cm\epsffile{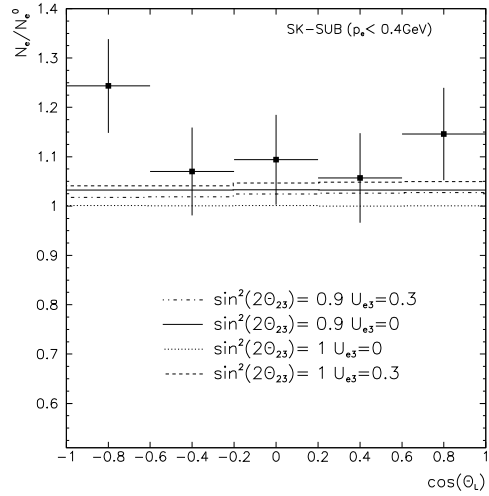}
\vspace{-1cm}
\caption{\it The zenith angle distributions of the $e-like$ events 
(ratio of number of events with, $N_{e}$,  and without, $N_{e}^0$, oscillations)
in the sub-GeV range ($p < 0.4$ GeV) for $\Delta m_{12} = 5 \cdot 10^{-5}$ eV$^2$ 
and different values of $\sin^2 2\theta_{23}$ and $|U_{e3}| \equiv s_{13}$. 
Dotted line - no $s_{13}$ effect, and completely screened effect of 1-2 mixing; 
solid line - effect of 1-2 mixing only  ($s_{13} = 0$);  
dashed line - the effects of interference for maximal 2-3 mixing (direct 1-2 contribution is screened); 
the dash-dotted line -  the effect of interference and 1 - 2 mixing. 
For presently favored value of $\Delta m_{12}^2$ all the effects which involve 1-2 mixing 
should be enhanced by factor 1.5. 
Also shown are the SuperKamiokande experimental points.}   
\label{zenith}
\end{center}
\end{figure}

If $s_{13} \neq 0$ the  interference 
of effects produced by the solar oscillation parameters and $s_{13}$   
should be  taken into account.  
In the sub-GeV range, the non-zero $s_{13}$ induces  the interference 
of the survival, $A_{ee}$, and transition, $A_{e\mu}$, amplitudes  
of the $2\nu$ system with $\Delta m^2_{12}$ and $\theta_{12}$.  
The relative change of the $\nu_e$ flux can be written as~\cite{PSint} 
\begin{eqnarray}
\frac{F_e}{F_e^0} - 1 = P_2 (r \cos^2 \theta_{23} - 1) - 
\nonumber\\
r s_{13} \sin2 \theta_{23} Re(A_{ee}^* A_{e\mu} ) - 
\nonumber\\ 
s_{13}^2 [2(1 - r \sin^2 \theta_{23})  + P_2 (r - 2)],  
\label{sub-int}
\end{eqnarray}
where $P_2 \equiv |A_{e\mu}|^2$. 
The interference effect  given 
by the second term in the RH side of (\ref{sub-int}) has the following properties: 
(i) it is linear in $s_{13}$,  (ii) has no screening factor, 
(iii)  has opposite signs for neutrinos and antineutrinos,  
(iv)  maximal for $\Delta m^2_{12} = 7 \cdot 10^{-5}$ eV$^2$. 
In fig.~\ref{zenith} from~\cite{PSint} we show the zenith 
angle distributions of the $e$-like events 
for different values of mixing. 

The interference term gives the dominant contribution 
to the excess of $e$-like events if 2-3 mixing is close to the maximal one. 
In maximum we find: 
\be
\epsilon^{int} \sim 0.16 s_{13}. 
\ee

\subsection{Sterile neutrinos}

The  $\nu_{\mu} - \nu_s$  oscillations are excluded as the dominant 
solution of the atmospheric neutrino problem. 
The data give also strong bound on  partial transition to the sterile component,   
$\nu_{\mu} \rightarrow   \cos \theta_s \nu_{\tau} + \sin \theta_s \nu_s$: 
$\sin^2 \theta_s < 0.2 - 0.3$ in the context of single $\Delta m^2$.

Existence of sterile neutrinos can lead to 
new manifestations in the context of more than one $\Delta m^2$. 
The fourth  neutrino  
with $\Delta m^2 \sim (0.5 - 1)$ eV$^2$ (motivated by LSND) 
has the MSW resonances in matter of the Earth  in the energy 
range (0.5 - 1.5) TeV \cite{br}. For the (3+1) scheme with normal mass hierarchy 
the resonances, and consequently, the resonance enhancement of oscillations,  
are in the ($\nu_e - \nu_s$) and ($\bar \nu_{\mu, \tau} - \bar \nu_s$) channels. 
The oscillations driven by $\Delta m^2_{LSND}$ lead to~\cite{br}

\begin{figure}[h!]
\begin{center}
\vspace{-0.5cm}
\hspace{6cm} \epsfxsize13cm\epsffile{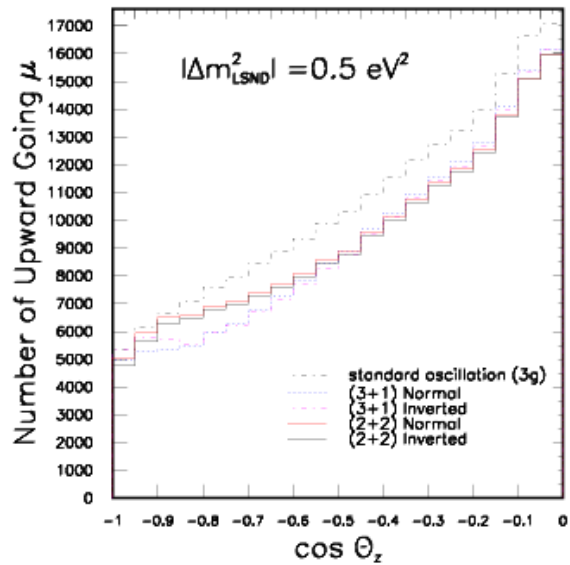}  
\vspace{-3.5cm}
\caption{\it 
The zenith angle distributions of the upward-going muons  
for different mass and mixing scenarios with large $\Delta m^2$.   
From $^{26}$.}
\label{ice}
\end{center}
\end{figure}

1). modification of the the zenith angle distribution of the 
upward-going muons (disappearance due to transition into sterile state),  

2). appearance of  fluxes of the electron neutrinos 
(strongly suppressed in the original flux at high energies) and 
tau neutrinos. These fluxes produce via the  CC interactions an additional 
contribution to the cascade events. 

For neutrinos crossing the core of the Earth  
the parametric enhancement of oscillations take place. 
The zenith angle distributions of events 
calculated for the Ice-Cube detector~\cite{br} are shown in the fig.~\ref{ice}.

\section{LSND: ultimate neutrino anomaly?}

Three different possibilities to reconcile the LSND signal 
with the solar and atmospheric neutrino results are under discussion:  
(i). Additional sterile neutrinos in the context of 
(3+1) or (3+2) schemes; 
(ii). Non-standard neutrino interactions; (iii). CPT violation. 

KamLAND  and some other recent results   
have changed  status of these possibilities.  

\subsection{(3 + 1) scheme}

\begin{figure}[h!]
\begin{center}
\vspace{0.1cm}
\hspace{-0.1cm} \epsfxsize5cm\epsffile{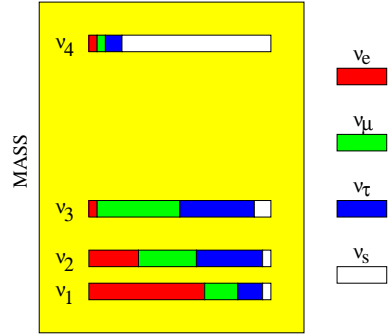}
\vspace{0.1cm}
\caption{\it The mass and flavor spectrum of the (3+1)-scheme.}
\label{three1}
\end{center}
\end{figure}

The main problem of the (3 + 1) scheme (fig.~\ref{three1}) is that the predicted LSND signal,  
which is consistent with the results  of other short base-line experiments  
(BUGEY, CHOOZ, CDHS, CCFR, KARMEN~\cite{KARMEN1}) as well as the atmospheric neutrino data,  is too 
small: 
the probability is about $3\sigma$ below the LSND measurement. 
Introduction of the second sterile  neutrino 
with $\Delta m^2 > 8$ eV$^2$ may help~\cite{PS31}. 
It was shown~\cite{sorel}  that  the second neutrino with  
$\Delta m^2 \sim 22$ eV$^2$ and specific mixing parameters 
can enhance the predicted LSND signal by (60 - 70) \% in comparison 
with (3 + 1) scheme.

However, the additional sterile neutrino aggravates the cosmological problems. 
This second (as well as the first one) 
sterile neutrino  equilibrates in the Early Universe 
unless significant ($> 10^{-5}$) lepton asymmetry existed in the epoch 
with $T \sim  10 - 20$ MeV and later. Equilibrium concentrations violate the   
nucleosynthesis, large scale structure bounds. 

Mild (factor of 2 - 3) improvement of the present CDHS bound on 
$\nu_{\mu}$-disappearance (by MiniBOONE ?) can exclude these possibilities~\cite{PS31}. 
Also improvements of the BUGEY  $\bar\nu_{e}$ - disappearance bound 
(probably by the new generation of the reactor experiments) can do similar job.

Notice that the (3 + 1) scheme is of interest even independently of the
LSND result.  The mass gap of the fourth (mainly sterile) neutrino with active
states can be arbitrary.  The existence of such a neutrino can be motivated
by the large lepton mixing.  Even very small coupling  of sterile neutrino
can modify the original $3 \times 3$ mass matrix with small flavor mixing  in
such a way that the flavor mixing will be enhanced~\cite{bal}.

On the other hand,  possible  existence of mixing with sterile neutrino leads to  
uncertainty in interpretation  of the results on neutrino masses and mixing 
unless the additional neutrinos will be discovered and their characteristics
measured.

\subsection{Non-standard neutrino interactions}

The  LSND signal could be due to the 
anomalous  decay of muon~\cite{BaP}: 
\be
\mu^+ \rightarrow \bar{\nu}_e \bar{\nu}_{i} e^+, ~~~ (i = e , \mu, \tau). 
\label{mode}
\ee
Violation of the lepton number by two units, $|\Delta L| = 2$,  allows to 
avoid 
stringent bound from  non-observation of the $\mu \rightarrow e e e$ mode.  
The decay (\ref{mode}) can be induced by the exchange of new neutral 
scalar boson ($M \sim 300 - 500$ GeV). As a result,  the Lorentz 
structure of the decay differs from the standard one: the Michel parameter 
equals $\rho = 0$. 

The problem of this interpretation is to reconcile ``LSND with KARMEN". 
Now one cannot play with difference of the  baselines and 
the situation is equivalent  to the averaged 
oscillation case (large $\Delta m^2$) where KARMEN gives stronger bound,  essentially 
excluding the LSND result.  The $\rho = 0$ feature  of new interaction does not help~\cite{KARMEN}. 

New experiment  TWIST at TRIUMPF~\cite{TWIST} will measure the Michel parameter 
with high accuracy which will allow us to check deviation  
from the standard value $\rho = 3/4$. 
  
\subsection{CPT-violation}

After KamLAND the ultimate possibility is  
the spectrum with 
$\Delta m^2_{sun}$ and $\Delta m^2_{atm}$ splittings in the neutrino 
channel and $\Delta m^2_{LSND}$ and $\Delta m^2_{KL}$ splittings  
in the antineutrino channel~\cite{stru,BLyk}. 

In this case, no oscillation effect should be seen by LSND
in the neutrino channel (decay in flight sample). In
fact, here the evidence of oscillations is at level $1\sigma$ only.

The main problem of the model is the  description of the atmospheric neutrino data   
(tension between the zenith angle distribution of the $\mu$-like events and the excess 
of the $e$-like events). 
In the antineutrino channel, the oscillations driven by 
$\Delta m^2_{LSND}$ are averaged and the effect due to $\Delta m^2_{KL}$  
is relatively weak. 
Furthermore, according to~\cite{BLyk}, the best fit 
corresponds to the non-maximal $\bar{\nu}_{\mu} - \bar{\nu}_{\tau}$ 
mixing. In this case the screening factor (\ref{sub-e}) is not small 
and one expects significant effect of the $\bar{\nu}_e$ oscillations  
driven by  the KamLAND oscillation parameters.  
A rough estimation gives  $(\sim 10 - 15)\%$ excess of the $e$-like events 
in the sub-GeV range.  

According this scheme, KamLAND does not check solar neutrino 
solution and therefore whole spectrum of possibilities (LOW, VO, SFP) 
is not yet excluded. For MiniBOONE one predicts null oscillation 
result in the neutrino channel, but positive signal in the antineutrino channel.

\section{Supernova Neutrinos}

The KamLAND result has important impact both on the 
interpretation of the signal from 
SN1987A and  on the program of future SN neutrino detection.  

``After KamLAND'' we can definitely say that  the 
effects of antineutrino flavor conversion 
have been observed already in 1987:  namely, effects of 

$\bullet$ $\bar{\nu}_e$  conversion  inside the star,  

$\bullet$ (probably) oscillations in the matter of the Earth; furthermore the  
oscillation effects  were different for Kamioka, IMB and Baksan  detectors.  

Specific effects depend on the type of mass hierarchy and value of $s_{13}$. 
In the case of  normal mass hierarchy the adiabatic $\bar{\nu}_e \rightarrow \bar \nu_1$  
and   $\bar{\nu}_{\mu, \tau} \rightarrow \bar \nu_2$ transitions 
occurred inside the star and then $\nu_1$ and $\nu_2$ oscillated inside 
the Earth~\cite{LS-87}.

With future SN burst detections one can 

(i) get information about $s_{13}$ 
(put upper or lover bound or measure it, depending on the 
true value of $s_{13}$), 

(ii) establish the mass hierarchy,  

(iii) test existence of sterile neutrinos.  

Main problem of  this program is that
the   original fluxes are not well known. Moreover, 
the neutrinos of all species are produced in the cooling phase 
which substantially diminishes the observable effects. (The effects  are  
proportional to the difference of original fluxes.) 

Identification and comparison of the neutrino and antineutrino signals is crucial.
However, small number of the expected  $\nu_e$- events  adds more uncertainties. 
So, developments of the high statistics $\nu_e$-detectors of SN neutrinos are highly welcomed.

In this connection, an  important task is to find
the ``star model-independent'' observables
which encode the information about conversion effects.

\subsection{LBL with supernova neutrinos}

Study of the Earth matter effects on the SN neutrinos 
is one possibility to get ``star model-independent'' information on neutrino parameters. 
In a sense one can perform the long baseline experiment with 
SN neutrinos. The beam uncertainties are  controlled if 
(i) two well separated  detectors are used,  
(ii) properties of medium  are known. 
Comparison of signals from the two detectors  allows one 
to establish  effect of oscillations inside the Earth. 

If $\sin^2 \theta_{13}  > 10^{-4}$, the  appearance 
of the Earth matter effect in $\bar \nu_e$  ($\nu_e$) channel 
will testify for the normal (inverted) mass hierarchy. 
Independently of $\sin^2 \theta_{13}$ 
value,  the  very fact of the absence of the Earth matter 
effect in the $\nu_e$  ($\bar \nu_e$) channel will exclude the inverted 
(normal) mass hierarchy~\cite{LSear}.  

Actually, the  existence of the Earth matter effect can be established with 
one detector: at high energies one predicts characteristic oscillatory 
distortion of the energy spectrum  which increases with 
energy~\cite{DS,LSear}. 

\subsection{Shock wave effect}

It was argued recently that the shock wave may reach the region 
of the neutrino conversion, $\rho \sim 10^{4}$ g/cc,  
after $t_s = (3 - 5)$ s from the bounce (beginning of the burst)~\cite{SF}.
Changing the density profile and therefore the adiabaticity,  
the shock front influences the conversion in 
the h-resonance characterized by the atmospheric 
$\Delta m^2_{13}$ and $\sin^2 \theta_{13}$,   
provided that $\sin^2 \theta_{13}  > 10^{-6}$. 

The following shock wave effects should be seen 
at some level in the neutrino (antineutrino) 
for normal (inverted) hierarchy:  

1). Change of the total number of events in time \cite{SF};  

2). Wave of softening of the spectrum which propagates 
in the energy scale  from  low energies to high energies 
\cite{Tak};  

3). Delayed Earth matter effect 
in the ``wrong" channel ({\it e.g.},  in neutrino channel for normal mass  hierarchy)  
\cite{LS03}.  

Modification of the density profile by the shock wave leads to 
appearance of additional resonances below the front. 
Effects of these resonance have been considered recently in 
\cite{lisi}.

Monitoring the shock wave with neutrinos is challenging but 
really exiting task which certainly deserves further 
considerations.  Studying the  shock wave effects on the properties 
of neutrino signals one can (in principle) get information on 
(i)  time of shock wave propagation, (ii) shock wave revival time, 
(iii)  velocity of propagation, (iv) density gradient in the front, 
(v) size of the front.  
This, in turn, can shed some light on the mechanism of
star explosions.

\section{Standard and Non-standard}

What is {\it standard} in the neutrino physics now: 

\begin{itemize}

\item
three neutrinos; 

\item
masses below $0.5 - 1$ eV; 

\item
bi-large or large-maximal mixing;  
 
\item
non-zero 1-3 mixing, probably  close to  the present upper bound;   

\item
smallness of neutrino mass related to 
the neutrality of neutrinos and their Majorana nature.  

\end{itemize}

What is beyond the standard picture? (i) new neutrino states (sterile neutrinos),  
(ii) new neutrino interactions; (iii) large anomalous magnetic moments, {\it etc.}. 
What is exotic? Effects of violation of the Lorentz invariance, CPT violation, equivalence 
principle, {\it etc.}. \\

\begin{figure}[h!]
\begin{center}
\hspace{-0.1cm} \epsfxsize8cm\epsffile{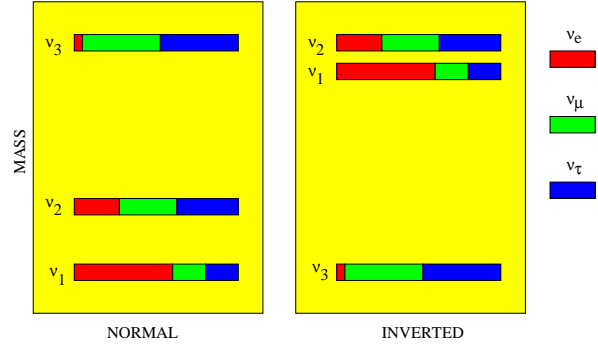} 
\caption{\it Neutrino mass and flavor spectra for the normal and inverted mass hierarchies.}
\label{sp}
\end{center}
\end{figure}

With the KamLAND result and resolution of the solar neutrino problem we made 
the next (after establishing  the oscillations in atmospheric neutrinos) 
major step in reconstruction of the neutrino mass and mixing 
spectrum: 

1). the  mass squared split of  $\nu_1$ and $\nu_{2}$ states, 
$\Delta m^2_{12}$, is determined;   

2).  distribution  of the electron flavor in $\nu_1$ and $\nu_{2}$ states  
 is measured (the best fit corresponds to 
 $|U_{e1}|^2 \approx 2 |U_{e2}|^2$);  

3) distribution of the muon and tau flavors in  $\nu_1$ and $\nu_{2}$ can be found  
with precision $O(s_{13})$ from the unitarity condition. 

These results are summarized in fig.~\ref{sp} which shows also  
what should be determined  to accomplish the picture: 

\begin{itemize}

\item
$U_{e3}$;

\item

type of mass hierarchy (in the case of hierarchical spectrum) 
or ordering of the states: normal, inverted; 

\item
type  of spectrum: hierarchical, non hierarchical, 
 partially degenerate, completely degenerate, which is equivalent to 
determination  of the absolute mass scale $m_1$. 

\end{itemize} 

What we cannot see in the plot is the CP-violating phases:  
Dirac phase $\delta$, and if neutrinos are Majorana particles,  
two Majorana phases. One needs also 
to establish the nature of neutrinos (Majorana-Dirac), or in general to measure their ``Majorana 
character".  
For a system of neutrinos one can introduce a parameter, related to  the lepton number 
whose change transforms pure Majorana neutrinos to quasi-Dirac and then to 
the Dirac neutrinos.

In this connection, the $\beta \beta_{0\nu}$ decay experiments are of the highest priority: 
the results will contribute (together with other measurements) to determination  of all unknown elements 
listed above.   

The KamLAND result and selection of the LMA solution has 
crucial consequences for the $\beta \beta_{0\nu}$ decay 
searches and interpretation of their results. 
Now we can say that  due to the large 1-2 mixing, there is a strong dependence 
of the effective Majorana mass of the electron neutrino, $m_{ee}$,  on the  
CP-violating phases.  This in turn, implies a possibility 
of substantial cancellation of contributions 
to  $m_{ee}$ from different mass eigenstates. 
Consequently, it is not  possible to 
determine the absolute scale of neutrino mass from 
the $\beta \beta_{0\nu}$ decay immediately.

\section{Conclusions}

In this paper I have discussed  topics on which the KamLAND result  
has an immediate impact. Furthermore, only  aspects related to the phenomenology 
of the neutrino mass and mixing have been covered. 
Even with these restrictions the review is far from being complete. 
I should mention here physics of the long baseline experiments  
elaborated largely before KamLAND. 
The first KamLAND result with confirmation of LMA 
has given further boost for realization of 
its experimental programs.

The main developments in neutrino physics during last 
30-35 years were related to various 
neutrino anomalies both real and fake.  
The review covered status of the anomalies after KamLAND with the 
questions: ``What is left" and   ``what are perspectives"? 

One can imagine several scenarios: 

$\bullet$ ``Standard scenario'' described in sect. 6: 
there is a  well defined program of reconstruction 
of the neutrino mass and flavor spectrum.  
It is characterized in terms of further tests,  precision measurements, 
searches for new physics. 

$\bullet$  Confirmation of the LSND result will open new perspectives
related to existence of new light neutral fermions, or CPT violation, 
or new interactions.

$\bullet$ New anomalies may appear  
which will lead to something unexpected 
(some hints from NuTeV,  $Z^0$-width measurements?). \\

``Without anomalies": we will work on the well defined program  
which consists of 

1). Determination of masses, mixings, CP-phases; precision measurements 
of parameters. Here, we face  ``technological problems'': 
determinations of the absolute mass scale and  CP-phases are indeed big challenge.

2). Searches for new physics beyond the standard picture, 
restrictions on exotics. The main issues are new neutrino interactions, 
new neutrino states, effects of violation of CPT, Lorentz invariance, 
equivalence principle, Pauli principle.

3). Identification of origins of the neutrino mass and mixing: 
that can include reconstruction of neutrino mass matrix, tests of the 
see-saw  mechanism and other possibilities (flavor violation processes, 
leptogenesis,  high energy experiments).

4). Applications of our knowledge of neutrino mass and mixing 
to Geophysics, Astrophysics, Cosmology. 

Notice that future high energy experiments  (LHC, TESLA ...) 
may have serious impact on this program. 

For details and further developments, see the talks at NOON2003~\cite{noon}.

\section*{Acknowledgments}

I am grateful to H. Minakata and Y. Suzuki for invitation
to give this talk and for hospitality during my stay in Kanazawa.


\end{document}